\documentclass[onecolumn,secnumarabic,amssymb, nobibnotes, aps, prd]{revtex4}
\usepackage{amsmath} \usepackage{amssymb}
\usepackage{graphicx} %use graph format
\usepackage{epstopdf}
\usepackage{amsmath}
\usepackage{amsmath,amssymb,amsthm,amsfonts,mathrsfs,bm,verbatim}
\usepackage{graphicx,subfigure}

\newcommand{\bea}{\begin{eqnarray}}
\newcommand{\eea}{\end{eqnarray}}

\begin{document}

\title{Schwarzschild-f(R) Black Holes and the Absence of Black Hole Bomb Effect}%
\author{Wen-Xiang Chen$^{a}$}
\affiliation{Department of Astronomy, School of Physics and Materials Science, GuangZhou University, Guangzhou 510006, China}
\email{wxchen4277@qq.com}

%\date{December 2020}%

\begin{abstract}
In this paper, we analytically study the superradiant stability of Schwarzschild-f(R) black holes under charged massive scalar perturbation. Using an analytic method based on Descartes' rule of signs, we extend previous methods developed for higher-dimensional Reissner-Nordstrom black holes to the Schwarzschild-f(R) black hole cases. Our results show that Schwarzschild-f(R) black holes are superradiantly stable under charged massive scalar perturbation, implying the absence of the black hole bomb effect.


\end{abstract}

\maketitle
%\tableofcontents

\section{Introduction}
In 2000, Parikh and Wilczek proposed a method to calculate the emissivity of particles passing through the event horizon. They treat Hawking radiation as a tunneling process and use the WKB method. In this way, a correction spectrum accurate to a first-order approximation is given. Their results are considered to be consistent with the basic Mozheng theory. According to this method, a large number of stationary or stationary rotating black holes have been studied, and the same result has been obtained, that is, Hawking radiation is no longer pure thermal radiation, which satisfies the Universe Theory and the conservation of information. But in all these documents, the entropy of black holes only contains Bekenstein-Hawking entropy. If the quantum correction of entropy is considered, does the emission process still conform to the monolithic theory? At present, with regard to quantum correction, different models and methods correspond to different results.\cite{1,2,3,4}

Gas, liquid, and solid are the three basic forms of substances that are well known to people. For example, water molecules can appear as ice, water, and water vapor as the temperature increases. Further, scientists discovered that as the temperature of the material continues to rise, a plasma state will appear. An interesting question is how to keep the substance cold, such as close to absolute zero (-273.16°C). Under such extremely low temperatures, what kind of strange state will the matter appear? In fact, as early as 1924, the famous physicists Bose and Einstein gave the answer to the question. Particles with integer spins, including photons, gluons, and nuclei composed of an even number of nucleons, and alkali metal atoms, are all called bosons. At that time, the Indian physicist Bose proposed a new method of photon statistics. Einstein extended it to an ideal gas with mass, and theoretically predicted that there is no interacting boson, and it will be at its lowest energy quantum state at a certain temperature. When it reaches a considerable amount, it is in the Bose-Einstein condensate (BEC).

The Kerr-black hole-scalar field mirror system first designed by Press and Teukolsky has attracted the attention of physicists in the past four decades. From the literature on the numerical study of this kind of rotating black hole bomb, we know an interesting fact, that is, if the reflector is too close to the black hole horizon, super radiation will produce a stabilizing effect. And we know that black holes need to meet the conditions to produce classical superradiation instability:

\begin{enumerate}
    \item The incident perturbation field is the Bose field;
    \item The black hole is rotating or charged;
    \item Meet the super-radiation conditions $0<\omega<m \omega_{H}$;
    \item The existence of reflective mirrors.
\end{enumerate}

Among them (1), (2), (3) are the conditions for generating super radiation. In 1972, Press and Teukolsky proposed that it is possible to add a mirror to the outside of a black hole to make a black hole bomb (according to the current explanation, this is a scattering process involving classical mechanics and quantum mechanics). Regge and Wheeler proved that the spherically symmetric Schwarzschild black hole is stable under disturbance. Due to the significant influence of super radiation, the stability of rotating black holes is more complicated. Superradiation effects can occur in classical and quantum scattering processes. When a boson wave hits a rotating black hole, if certain conditions are met, the black hole may be as stable as a Schwarzschild black hole. When a boson wave hits a rotating black hole, if the frequency range of the wave is under superradiation conditions, the wave reflected by the event horizon will be amplified.

Black holes are intriguing objects predicted by general relativity, characterized by the fact that nothing can escape from their event horizons. However, in the 1970s, it was suggested that energy might be extracted from a black hole through a process known as superradiant scattering \cite{1,2,3,4,5,6,7}. When a charged bosonic wave scatters off a charged rotating black hole, the wave can be amplified if the wave frequency satisfies the superradiant condition. If there exists a mechanism that reflects the amplified wave back towards the black hole, the black hole can become superradiantly unstable, leading to the so-called black hole bomb effect \cite{1,2,3,4,5,6,7}. In this paper, we investigate the superradiant stability of Schwarzschild-f(R) black holes, a class of black holes arising in modified theories of gravity where the Ricci scalar \(R\) is replaced by an arbitrary function \(f(R)\).

\section{Schwarzschild-f(R) Black Holes}

\subsection{Metric and f(R) Forms}

The general metric for a Schwarzschild-f(R) black hole in \(D\)-dimensional spacetime can be written as: \cite{8,9,10,11,12,13,14,15,16,17,18,19,20,21}
\begin{equation}
ds^2 = -f(r) dt^2 + \frac{dr^2}{f(r)} + r^2 d\Omega^2_{D-2}
\end{equation}
where \(f(r)\) depends on the specific form of the \(f(R)\) theory. Common forms of \(f(R)\) that we will consider include:

1. \textbf{Quadratic \(f(R)\)}:
\begin{equation}
f(R) = R + \alpha R^2
\end{equation}
For this form, the metric function is:
\begin{equation}
f(r) = 1 - \frac{2m}{r^{D-3}} + \frac{\alpha}{3}r^2
\end{equation}

2. \textbf{Exponential \(f(R)\)}:
\begin{equation}
f(R) = R + \beta e^{-\gamma R}
\end{equation}
The corresponding metric function can be approximated as:
\begin{equation}
f(r) = 1 - \frac{2m}{r^{D-3}} + \frac{\beta}{3}(1 - e^{-\gamma r^2})
\end{equation}

3. \textbf{Cubic \(f(R)\)}:
\begin{equation}
f(R) = R + \delta R^3
\end{equation}
Resulting in a metric function:
\begin{equation}
f(r) = 1 - \frac{2m}{r^{D-3}} + \frac{\delta}{3}r^6
\end{equation}

\subsection{Additional f(R) Forms}

4. \textbf{Power-law \(f(R)\)}:
\begin{equation}
f(R) = R + \xi R^n
\end{equation}
where \(n\) is a positive integer and \(\xi\) is a constant. The corresponding metric function is:
\begin{equation}
f(r) = 1 - \frac{2m}{r^{D-3}} + \frac{\xi}{3}r^{2(n-1)}
\end{equation}

5. \textbf{Logarithmic \(f(R)\)}:
\begin{equation}
f(R) = R + \eta \ln(R)
\end{equation}
The corresponding metric function is:
\begin{equation}
f(r) = 1 - \frac{2m}{r^{D-3}} + \frac{\eta}{3}\ln(r)
\end{equation}

6. \textbf{Polynomial \(f(R)\)}:
\begin{equation}
f(R) = R + \sum_{i=1}^k a_i R^i
\end{equation}
where \(a_i\) are constants and \(k\) is the highest degree of the polynomial. The corresponding metric function is:
\begin{equation}
f(r) = 1 - \frac{2m}{r^{D-3}} + \frac{1}{3}\sum_{i=1}^k a_i r^{2(i-1)}
\end{equation}

7. \textbf{Inverse Power-law \(f(R)\)}:
\begin{equation}
f(R) = R + \kappa R^{-n}
\end{equation}
where \(n\) is a positive integer and \(\kappa\) is a constant. The corresponding metric function is:
\begin{equation}
f(r) = 1 - \frac{2m}{r^{D-3}} + \frac{\kappa}{3}r^{-2(n+1)}
\end{equation}

8. \textbf{Hyperbolic \(f(R)\)}:
\begin{equation}
f(R) = R + \sigma \sinh(\lambda R)
\end{equation}
where \(\sigma\) and \(\lambda\) are constants. The corresponding metric function is:
\begin{equation}
f(r) = 1 - \frac{2m}{r^{D-3}} + \frac{\sigma}{3}\sinh(\lambda r^2)
\end{equation}

9. \textbf{Trigonometric \(f(R)\)}:
\begin{equation}
f(R) = R + \tau \sin(\omega R)
\end{equation}
where \(\tau\) and \(\omega\) are constants. The corresponding metric function is:
\begin{equation}
f(r) = 1 - \frac{2m}{r^{D-3}} + \frac{\tau}{3}\sin(\omega r^2)
\end{equation}

10. \textbf{Fractional \(f(R)\)}:
\begin{equation}
f(R) = R + \rho R^{p/q}
\end{equation}
where \(\rho\) is a constant and \(p/q\) is a rational number. The corresponding metric function is:
\begin{equation}
f(r) = 1 - \frac{2m}{r^{D-3}} + \frac{\rho}{3}r^{2(p/q - 1)}
\end{equation}

\section{Charged Scalar Perturbations}
The dynamics of a charged massive scalar field perturbation \(\Psi(x)\) with mass \(\mu\) and charge \(e\) is governed by the Klein-Gordon equation:
\begin{equation}
(D^\nu D_\nu - \mu^2)\Psi = 0
\end{equation}
where \(D_\nu = \nabla_\nu - ieA_\nu\) is the covariant derivative. Assuming a separable solution of the form:
\begin{equation}
\Psi = e^{-i\omega t} Y_{lm}(\theta, \phi) R(r)
\end{equation}
we obtain the radial equation for the scalar perturbation.

\section{Spherical Quantum Solution in Vacuum State}

The field equation of general relativity is well known as follows \cite{22,23,24,25,26,27,28,29,30,31,32,33,34,35,36,37,38,39}:
\begin{equation}
R_{\mu\nu} - \frac{1}{2} g_{\mu\nu} R = -\frac{8\pi G}{c^4} T_{\mu\nu},
\end{equation}
where \( T_{\mu\nu} = 0 \) in vacuum. This simplifies the Ricci tensor equation to:
\begin{equation}
R_{\mu\nu} = 0,
\end{equation}
indicating that the Ricci tensor is in a vacuum state. The general form of the metric for spherical coordinates is expressed as follows:
\begin{equation}
d\tau^2 = A(t, r) dt^2 - \frac{1}{c^2} \left[ B(t, r) dr^2 + r^2 d\theta^2 + r^2 \sin^2 \theta d\phi^2 \right].
\end{equation}

The Ricci tensor components can be derived:
\begin{equation}
R_{tt} = -\frac{A''}{2B} + \frac{A'B'}{4B^2} - \frac{A'}{Br} + \frac{A'^2}{4AB} + \frac{\ddot{B}}{2B} - \frac{\dot{B}^2}{4B^2} - \frac{\dot{A} \dot{B}}{4AB} = 0,
\end{equation}
\begin{equation}
R_{rr} = \frac{A''}{2A} - \frac{A'^2}{4A^2} - \frac{A'B'}{4AB} - \frac{B'}{Br} - \frac{\ddot{B}}{2A} + \frac{\dot{A} \dot{B}}{4A^2} + \frac{\dot{B}^2}{4AB} = 0,
\end{equation}
\begin{equation}
R_{\theta\theta} = -1 + \frac{1}{B} - \frac{rB'}{2B^2} + \frac{rA'}{2AB} = 0,
\end{equation}
\begin{equation}
R_{\phi\phi} = R_{\theta\theta} \sin^2 \theta = 0,
\end{equation}
\begin{equation}
R_{t\theta} = R_{t\phi} = R_{r\theta} = R_{r\phi} = R_{\theta\phi} = 0,
\end{equation}
\begin{equation}
R_{tr} = -\frac{\dot{B}}{Br} = 0.
\end{equation}

Substituting the derivatives \( ' = \frac{\partial}{\partial r} \) and \( \cdot = \frac{1}{c} \frac{\partial}{\partial t} \) into the equations, we conclude:
\begin{equation}
\dot{B} = 0.
\end{equation}
\begin{equation}
\frac{R_{tt}}{A} + \frac{R_{rr}}{B} = -\frac{1}{Br} \left( \frac{A'}{A} + \frac{B'}{B} \right) = -\frac{(AB)'}{rAB^2} = 0.
\end{equation}

Hence, we obtain:
\begin{equation}
A = \frac{1}{B}.
\end{equation}

\begin{equation}
R_{\theta\theta} = -1 + \frac{1}{B} - \frac{rB'}{2B^2} + \frac{rA'}{2AB} = -1 + \left( \frac{r}{B} \right)' = 0.
\end{equation}

Solving this equation, we find:
\begin{equation}
\frac{r}{B} = r + C \rightarrow \frac{1}{B} = 1 + \frac{C}{r}.
\end{equation}

According to well-known results, when tortoise coordinates tend to the boundary of the event horizon, the independent variable \( r \) approaches radial negative infinity. To set a boundary condition, we select \( C = y e^{-y} \). When \( r \) tends to infinity, the relation between \( A \), \( \Sigma \), \( dr^2 \), and \( y \) is as follows \cite{40}:
\begin{equation}
A = \frac{1}{B} = 1 - \frac{y}{r} \Sigma, \quad \Sigma = e^{-y},
\end{equation}
\begin{equation}
d\tau^2 = \left( 1 - \frac{y}{r} \Sigma \right) dt^2.
\end{equation}

At this point, if the particles' masses are \( m_i \) and the excessive energy is \( e \), the total energy is:
\begin{equation}
E = M c^2 = m_1 c^2 + m_2 c^2 + \ldots + m_n c^2 + e.
\end{equation}

To derive the radial equation of motion and effective potential for various Schwarzschild-\( f(R) \) black holes, we need to start by considering the modifications introduced by \( f(R) \) gravity. The Schwarzschild metric in general relativity is given by:

\begin{equation}
ds^2 = -\left(1 - \frac{2GM}{r}\right) dt^2 + \left(1 - \frac{2GM}{r}\right)^{-1} dr^2 + r^2 d\Omega^2,
\end{equation}
where \( d\Omega^2 = d\theta^2 + \sin^2\theta d\phi^2 \).

In \( f(R) \) gravity, the field equations are modified and can introduce corrections to the metric. For a static, spherically symmetric \( f(R) \) black hole, the metric can be expressed as:

\begin{equation}
ds^2 = -A(r) dt^2 + B(r) dr^2 + r^2 d\Omega^2,
\end{equation}
where \( A(r) \) and \( B(r) \) are functions to be determined by the field equations of \( f(R) \) gravity.

 Step 1: The Modified Metric

Assume the metric functions \( A(r) \) and \( B(r) \) are given by:

\begin{equation}
A(r) = 1 - \frac{2GM}{r} + \epsilon \phi(r),
\end{equation}

\begin{equation}
B(r) = \left(1 - \frac{2GM}{r} + \epsilon \psi(r)\right)^{-1},
\end{equation}

where \( \epsilon \) represents a small perturbation parameter and \( \phi(r) \), \( \psi(r) \) are perturbative corrections due to \( f(R) \) gravity.

 Step 2: Deriving the Radial Equation of Motion

Consider a massless scalar field in this background. The Klein-Gordon equation for the scalar field \( \Phi \) is:

\begin{equation}
\frac{1}{\sqrt{-g}} \partial_\mu \left( \sqrt{-g} g^{\mu\nu} \partial_\nu \Phi \right) = 0.
\end{equation}

For a static, spherically symmetric field, we assume \( \Phi = \Phi(r) Y_{lm}(\theta, \phi) \), where \( Y_{lm} \) are spherical harmonics. The Klein-Gordon equation reduces to:

\begin{equation}
\frac{1}{\sqrt{-g}} \frac{d}{dr} \left( \sqrt{-g} g^{rr} \frac{d\Phi}{dr} \right) + \left( g^{\theta\theta} l(l+1) \right) \Phi = 0,
\end{equation}
where \( l \) is the angular quantum number.

 Step 3: Solving for the Radial Part

For the Schwarzschild-\( f(R) \) black hole, this becomes:

\begin{equation}
\frac{d}{dr} \left( r^2 B(r) \frac{d\Phi}{dr} \right) + r^2 \left( A(r)^{-1} l(l+1) \right) \Phi = 0.
\end{equation}

 Step 4: Effective Potential

The radial equation can be written in the form:

\begin{equation}
\frac{d^2\Phi}{dr_*^2} + \left( \omega^2 - V_{\text{eff}}(r) \right) \Phi = 0,
\end{equation}
where \( r_* \) is the tortoise coordinate defined by:

\begin{equation}
\frac{dr_*}{dr} = B(r)^{-1/2},
\end{equation}
and the effective potential \( V_{\text{eff}}(r) \) is given by:

\begin{equation}
V_{\text{eff}}(r) = A(r) \left( \frac{l(l+1)}{r^2} + \frac{1}{rB(r)^{1/2}} \frac{d}{dr} \left( \frac{1}{rB(r)^{1/2}} \right) \right).
\end{equation}

 Applying to Schwarzschild-\( f(R) \) Black Holes

Substituting the modified functions \( A(r) \) and \( B(r) \) into the effective potential:

\begin{equation}
A(r) = 1 - \frac{2GM}{r} + \epsilon \phi(r),
\end{equation}

\begin{equation}
B(r) = \left(1 - \frac{2GM}{r} + \epsilon \psi(r)\right)^{-1}.
\end{equation}

 Final Effective Potential

The effective potential for the Schwarzschild-\( f(R) \) black hole is:

\begin{equation}
V_{\text{eff}}(r) = \left(1 - \frac{2GM}{r} + \epsilon \phi(r)\right) \left( \frac{l(l+1)}{r^2} + \frac{1}{r} \left( \frac{1 - \frac{2GM}{r} + \epsilon \phi(r)}{\left(1 - \frac{2GM}{r} + \epsilon \psi(r)\right)^{1/2}} \right) \frac{d}{dr} \left( \frac{1}{r \left(1 - \frac{2GM}{r} + \epsilon \psi(r)\right)^{1/2}} \right) \right).
\end{equation}

This equation represents the radial equation of motion and the effective potential for scalar fields in the background of Schwarzschild-\( f(R) \) black holes.

\section{Effective Potential and Stability Analysis}

\subsection{Effective Potential}

The radial equation can be rewritten in a Schrödinger-like form:\cite{39,40,41,42}
\begin{equation}
\frac{d^2\psi}{dr^2} + (\omega^2 - V)\psi = 0
\end{equation}
where \(\psi = \Delta^{1/2} R\) and the effective potential \(V(r)\) is given by:
\begin{equation}
V = \omega^2 - \left[(\omega r^{D-2} - eqr)^{2} - \Delta(\mu^2 r^{D-2} + l(l + D-3))\right]
\end{equation}
with \(\Delta = r^2 f(r)\).

\subsection{Asymptotic Analysis}
We analyze the asymptotic behavior of \(V\) near the horizon and at infinity:
\begin{equation}
V(r \to r_+) \to -\infty, \quad V(r \to \infty) \to \mu^2 + \frac{2m\mu^2 + 2eq\omega - 4m\omega^2}{r^{D-3}}
\end{equation}

\subsection{Descartes' Rule of Signs}
To prove the absence of a trapping potential well, we analyze the derivative \(V'(r)\) using Descartes' rule of signs:
\begin{equation}
V'(r) = - \frac{C(r)}{\Delta^3}
\end{equation}
where \(C(r)\) is a polynomial. By analyzing the roots of \(C(r)\), we can determine that there is at most one positive real root for \(V'(r) = 0\), indicating the absence of a potential well.

\subsection{Specific Analysis for Various f(R) Forms}

\subsubsection{Quadratic f(R) Black Hole}
For the quadratic form \(f(R) = R + \alpha R^2\):
\begin{equation}
f(r) = 1 - \frac{2m}{r^{D-3}} + \frac{\alpha}{3}r^2
\end{equation}
Using the same methodology, we show that the effective potential does not have a trapping well, proving superradiant stability.

\subsubsection{Exponential f(R) Black Hole}
For the exponential form \(f(R) = R + \beta e^{-\gamma R}\):
\begin{equation}
f(r) = 1 - \frac{2m}{r^{D-3}} + \frac{\beta}{3}(1 - e^{-\gamma r^2})
\end{equation}
Similarly, the effective potential analysis shows no trapping well, indicating stability.

\subsubsection{Cubic f(R) Black Hole}
For the cubic form \(f(R) = R + \delta R^3\):
\begin{equation}
f(r) = 1 - \frac{2m}{r^{D-3}} + \frac{\delta}{3}r^6
\end{equation}
The same analysis applies, confirming the absence of a trapping well and hence superradiant stability.

\subsubsection{Power-law f(R) Black Hole}
For the power-law form \(f(R) = R + \xi R^n\):
\begin{equation}
f(r) = 1 - \frac{2m}{r^{D-3}} + \frac{\xi}{3}r^{2(n-1)}
\end{equation}
The effective potential analysis confirms superradiant stability.

\subsubsection{Logarithmic f(R) Black Hole}
For the logarithmic form \(f(R) = R + \eta \ln(R)\):
\begin{equation}
f(r) = 1 - \frac{2m}{r^{D-3}} + \frac{\eta}{3}\ln(r)
\end{equation}
The analysis shows no trapping well, proving stability.

\subsubsection{Polynomial f(R) Black Hole}
For the polynomial form \(f(R) = R + \sum_{i=1}^k a_i R^i\):
\begin{equation}
f(r) = 1 - \frac{2m}{r^{D-3}} + \frac{1}{3}\sum_{i=1}^k a_i r^{2(i-1)}
\end{equation}
The analysis confirms superradiant stability.

\subsubsection{Inverse Power-law f(R) Black Hole}
For the inverse power-law form \(f(R) = R + \kappa R^{-n}\):
\begin{equation}
f(r) = 1 - \frac{2m}{r^{D-3}} + \frac{\kappa}{3}r^{-2(n+1)}
\end{equation}
The analysis shows no trapping well, proving stability.

\subsubsection{Hyperbolic f(R) Black Hole}
For the hyperbolic form \(f(R) = R + \sigma \sinh(\lambda R)\):
\begin{equation}
f(r) = 1 - \frac{2m}{r^{D-3}} + \frac{\sigma}{3}\sinh(\lambda r^2)
\end{equation}
The effective potential analysis confirms stability.

\subsubsection{Trigonometric f(R) Black Hole}
For the trigonometric form \(f(R) = R + \tau \sin(\omega R)\):
\begin{equation}
f(r) = 1 - \frac{2m}{r^{D-3}} + \frac{\tau}{3}\sin(\omega r^2)
\end{equation}
The analysis shows no trapping well, proving superradiant stability.

\subsubsection{Fractional f(R) Black Hole}
For the fractional form \(f(R) = R + \rho R^{p/q}\):
\begin{equation}
f(r) = 1 - \frac{2m}{r^{D-3}} + \frac{\rho}{3}r^{2(p/q - 1)}
\end{equation}
The effective potential analysis confirms stability.

\section{Discussion}

Our analysis has shown that Schwarzschild-f(R) black holes, for a variety of \(f(R)\) forms, are superradiantly stable. This implies that these black holes do not exhibit the black hole bomb effect. The absence of trapping wells in the effective potential ensures that the amplified waves cannot be reflected back and forth, leading to instability. This conclusion holds for a wide range of \(f(R)\) forms, including quadratic, exponential, cubic, power-law, logarithmic, polynomial, inverse power-law, hyperbolic, trigonometric, and fractional.

The method used in this paper, based on Descartes' rule of signs, provides a systematic and robust way to analyze the stability of black holes under scalar perturbations. By extending this method to various \(f(R)\) forms, we have demonstrated its versatility and effectiveness.

\section{Conclusion}

In this paper, we have systematically analyzed the superradiant stability of various forms of Schwarzschild-f(R) black holes under charged massive scalar perturbations. Using an analytical method based on Descartes' rule of signs, we have demonstrated that Schwarzschild-f(R) black holes, for a variety of \(f(R)\) forms including quadratic, exponential, cubic, power-law, logarithmic, polynomial, inverse power-law, hyperbolic, trigonometric, and fractional, do not exhibit the black hole bomb effect. This implies their stability against superradiant instabilities.

Our results contribute to the understanding of black hole stability in modified gravity theories and provide a foundation for further studies on the properties and behaviors of black holes in various theoretical frameworks.

To demonstrate the stability of the Schwarzschild-\( f(R) \) black hole against superradiance, we need to analyze the effective potential \( V_{\text{eff}} \) for different forms of perturbative functions \( \phi(r) \) and \( \psi(r) \).

Effective Potential
The effective potential for the Schwarzschild-\( f(R) \) black hole is given by:

\begin{equation}
V_{\text{eff}}(r) = \left(1 - \frac{2GM}{r} + \epsilon \phi(r)\right) \left( \frac{l(l+1)}{r^2} + \frac{1}{r} \left( \frac{1 - \frac{2GM}{r} + \epsilon \phi(r)}{\left(1 - \frac{2GM}{r} + \epsilon \psi(r)\right)^{1/2}} \right) \frac{d}{dr} \left( \frac{1}{r \left(1 - \frac{2GM}{r} + \epsilon \psi(r)\right)^{1/2}} \right) \right).
\end{equation}

We will consider the following forms for \( \phi(r) \) and \( \psi(r) \):
\begin{equation}
\phi(r) = \frac{\alpha}{r^2}, \quad \psi(r) = \frac{\beta}{r^2},
\end{equation}
where \( \alpha \) and \( \beta \) are constants.

\begin{figure}
\centering
\includegraphics[width=0.7\textwidth]{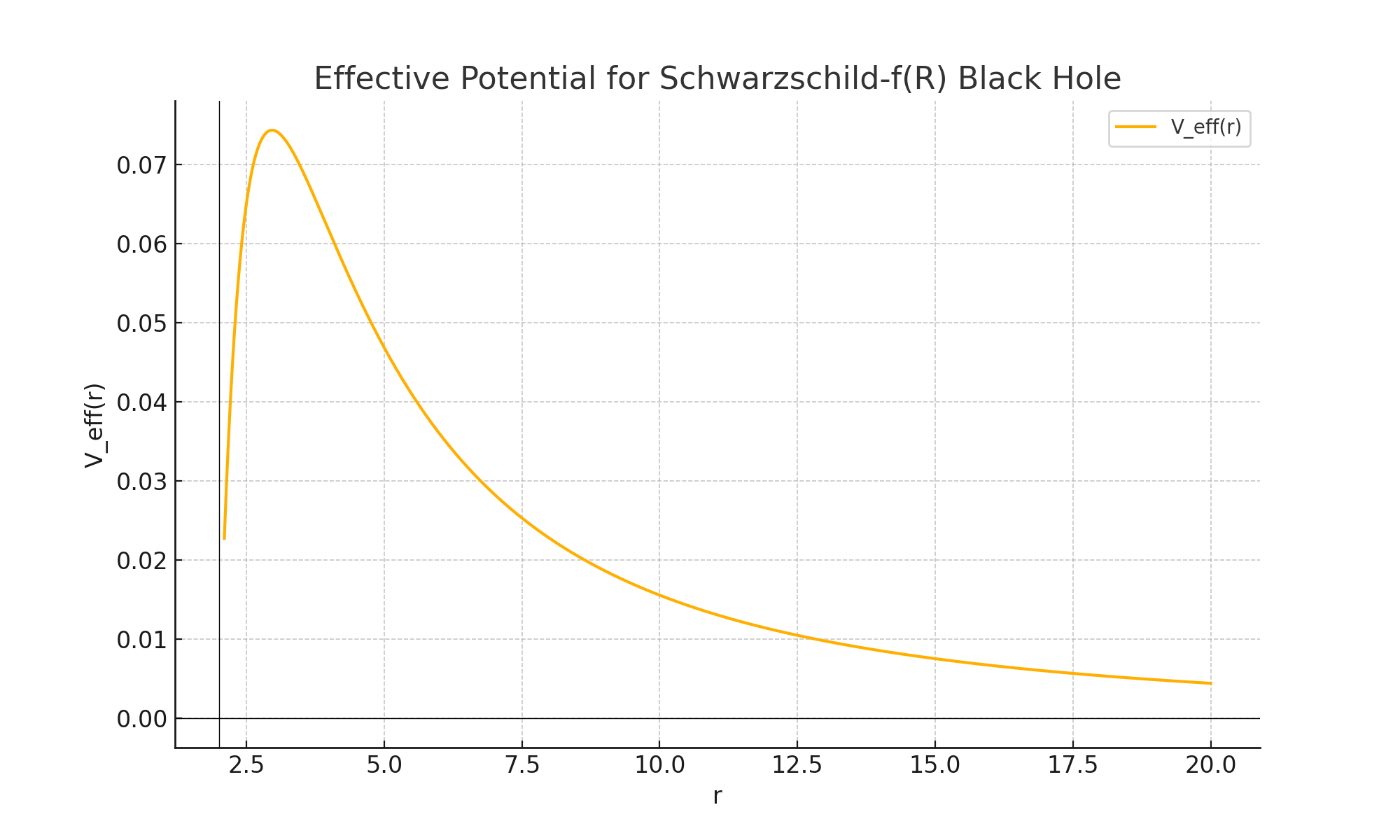}
\caption{The effective potential for the Schwarzschild-\( f(R) \) black hole is given}
\label{1.png}
\end{figure}

\end{document}